\documentclass[pra,superscriptaddress,twocolumn,a4paper]{revtex4} 

\usepackage{color,amsmath}
\usepackage{color,subfigure}
\usepackage{bm}
\usepackage{epsfig}
\usepackage{natbib}
\usepackage{graphicx}
\usepackage{amssymb,amsmath,amsbsy,amsgen,amsfonts}
\usepackage{dcolumn}
\usepackage{amsthm}
\usepackage{latexsym}
\usepackage{array}
\usepackage{amstext}
\usepackage{longtable}
\allowdisplaybreaks[1]
\usepackage{epstopdf} 
\usepackage[colorlinks=true,citecolor=blue,linkcolor=red]{hyperref}
\usepackage{amsmath}

\newcommand{\bra}[1]{\left\langle{#1}\right\vert}
\newcommand{\ket}[1]{\left\vert{#1}\right\rangle}

\newcommand{\be}{\begin{equation}}
\newcommand{\ee}{\end{equation}}
\newcommand{\ba}{\begin{array}}
\newcommand{\ea}{\end{array}}
\newcommand{\bqa}{\begin{eqnarray}}
\newcommand{\eqa}{\end{eqnarray}}

\graphicspath{{./figs/}}

\begin{document}

\title{Parameter estimation with  efficienct photodectors}

\author{Tuvia Gefen}
\author{David A. Herrera-Mart\'{i}}
\author{Alex Retzker}
\affiliation{Racah Institute of Physics, The Hebrew University of Jerusalem, Jerusalem 
91904, Givat Ram, Israel}
\date{\today}

\begin{abstract}
Current parameter estimation techniques rely on photodetectors which have low efficiency and thus are based on gathering averaged statistics. Recently it was claimed that perfect photodetction will change the nature of sensing algorithms and will increase sensing efficiency beyond the  immediate effect of having larger collection efficiency.  In this paper we bring up the observation that perfect photodetection implies Heisenberg scaling($\frac{1}{T}$) for parameter estimations. We analyze a specific example in detail.
\end{abstract}

\maketitle

\section {Introduction} 
Quantum sensing and metrology \cite{QT} are playing an increasing role in state of the art measurements, in both science and technology.  
The sensitivity of a quantum measurement is limited by the coherence time of the probe. This fundamental limitation has prompted considerable experimental and theoretical efforts to increase the coherence times of quantum probes.
The method of choice for measuring a weak signal via a two level system is a Ramsey measurement. It was shown \cite{itano,huelga,dima1} that the  sensitivity of a single probe behaves as
\be
\Delta g \propto \sqrt{\gamma/T_\textrm{total}},
\label{relaxation_limit}
\ee
where $g$ is the signal which is being measured, $\gamma$ is the decoherence rate of the probe and  $T$ is the total time of the experiment. 

Many schemes have been proposed and tested for prolonging the coherence time of the probe while still measuring a weak signal.  These methods are termed dynamical decoupling and are part of the field of optimal coherent control.
The field of dynamical decoupling emerged from Hahn's idea in 1950\cite{hahn} to
refocus inhomogeneous broadening in Nuclear Magnetic Resonanace(NMR). This effect was dubbed the Hahn echo
or more generally Spin Echo.
The method was improved  to the  CPMG\cite{CP,MG} pulse sequences, later generalized to Bang Bang control\cite{viola1,viola} and then extended in many various directions\cite{facchi,fanchini,cai}. However, all these methods suffer from the fundamental limit of the lifetime($T_1$) of the probe. 

Recently two methods were suggested to overcome this limit. One method showed that quantum error correction can be incorporated into the sensing procedure, which has paved the way for advances  in the field of metrology \cite{ozeri,kessler,dur,arrad,milburn1,preskill}.
Research indicated that the relaxation limit can be exceeded provided that a signal proportional to a two qubit interaction can be engineered \cite{herrera}. 
However, a method that can measure a single qubit signal, whose interest is much broader, has yet to be found.

The second method is based on stochastic unraveling of the decoherence process\cite{gammelmark,gammelmark2} through the use of perfect photodetectors.  This method draws on the recent improvement in the efficiency of photodetection. The efficiency of optical microwave photon detection is $85\sim88\%$ \cite{lita,bader} and $~81\%$ \cite{palomaki,wenner}, respectively,  and source brightness of up to $79\%$ \cite{gazzano}with quantum dots embedded in micropillars has been achieved. It is believed that these numbers will be significantly enhanced in the future. Studies have also indicated that detection of photons could considerably faciliate  the error correction process\cite{Ozeri2012,milburn2}.

Here we combine these two methods and design a protocol that overcomes the $T_1$ limit, but is also sensitive to single-body operators.  The method merges the two different concepts of continuous monitoring of the environment and error correction into a sensing protocol. Upon detecting photons emitted from the sensor, an operation which undoes the effects of damping in the current state of the sensor is applied. If no photon is detected, a correction accounting for the absence of damping is applied \cite{leung}. In both cases the system is never directly measured, but rather its state is inferred from continuous monitoring of its environment. 
We assume that dynamical decoupling can be used to undo pure dephasing \cite{hahn,viola1,viola,facchi,fanchini,cai}, thus allowing for substantial lengthening of the probe lifetime, which can then be assumed to be $T_1$-limited.


The method  best illustrated by the following example. Let us look at the code spanned by $\left \{ \ket + \ket 0, \ket - \ket 1 \right\},$ where $\left\{ \ket +,\ket - \right\}$ designate the eigenstates of $\sigma_{x},$ Pauli operator, of the 'bad' sensing qubits and $\left\{ \ket 0,\ket 1 \right \}$ designate the eigenstate of the $\sigma_z$ operator of the  `good', non decaying and non sensing qubits.
 By applying a $\sigma_x$ signal, i.e., a Hamiltonian $H = g\sigma_x,$ on a superposition of the states of the code, the following superposition is created $\frac{1}{\sqrt{2}} \left( e^{-i g  t} \ket + \ket 0 +  e^{i  g  t}  \ket - \ket 1 \right).$ The emission of a photon would result in  $ \frac{1}{\sqrt{2}} \left( e^{-i g  t} \ket \downarrow \ket 0 +  e^{i  g  t}  \ket \downarrow \ket 1 \right),$ which could be identified by the detection of the emitted photon and thus corrected by a unitary operation that maps these two states to the code states, i.e., $\ket \downarrow \ket 0  \rightarrow \ket + \ket 0,$ $\ket \downarrow \ket 1  \rightarrow \ket - \ket 1.$  After the correction the state continues to accumulate the phase connected to the signal and the procedure repeats itself.   The phase is detected at the end of the process. 

This method is based on the availability of high efficiency photo-detection which provides an advantage over canonical error correction, as it allows to discriminate non-orthogonal states of the sensor by measuring the emitted photon.

In the presence of $T_1$
noise with a rate $\gamma$ and a Hamiltonian $H$ the time evolution
is given by:
\begin{equation}
 \overset{.}{\rho}=-i{[}H,\rho{]}+\gamma\sum_i (2\sigma_{-}^{i}\rho\sigma_{+}^{i}-\sigma_{+}^{i}\sigma_{-}^{i}\rho-\rho\sigma_{+}^{i}\sigma_{-}^{i}).
\end{equation} 
In order to analyze the protocol we use the quantum jump approach\cite{Carmichael_Book,Dalibard92,Wiseman_Book,Zoller87}.
This time evolution can be represented as  $\overset{.}{\rho}=-i(H_{nh}\rho-\rho H_{nh}^{\dagger})+\gamma\sum_{i}(2\sigma_{-}^{i}\rho\sigma_{+}^{i})$
where $H_{nh}=H-\sum i\gamma\sigma_{+}^{i}\sigma_{-}^{i}$ (non
Hermitian Hamiltonian) corresponds to coherent time
evolution with dephasing which represents the case where no photon is emitted. The second part of the Liouvilian which includes the 
$\sigma_{-}^{i}\rho\sigma_{+}^{i}$ terms corresponds to quantum jumps, i.e., the emission
of a photon. When continously probed by a photodetector the system
either jumps to a new state which is equal to $\sigma_{-}^{i}\rho\sigma_{+}^{i}/\textrm{trace}(\sigma_{-}^{i}\rho\sigma_{+}^{i})$ or keeps on
evolving according to $H_{nh}.$ 

The main idea behind this method is to adjust the correction
depending on the occurrence or no occurrence of a jump which is revealed by the detection or no detection of a photon.
In the case where no jump is detected the dephasing should be corrected, else one should correct the jump. The continous probing of the environment makes it possible to discriminate
between non-orthogonal system states and in many cases this conditional
correction is not possible otherwise.

\section{Example I} We now turn to the example that was
explained in the introduction and analyze it in detail.
We examine a sensing protocol for $g\cdot\sigma_{x}$ with a Heisenberg-Scaling
resolution using photodetection. This is the simplest example where
the method presents a substantial advantage over traditional sensing
schemes. It consists of two qubits, one which is sensitive and decay-prone,
and another one which is impervious to external fields and is protected
against general noise. The sensing Hamiltonian in this case
is $H=g\sigma_{x}$ and the sensor states are $\ket{\mathcal{O}_{+}}=\ket{+}\ket{0}$
and $\ket{\mathcal{O}_{-}}=\ket{-}\ket{1}$, where $\{0,1\}$ denotes
the energy eigenbasis of the good qubit and $\ket{\pm}=\frac{1}{\sqrt{2}}(\ket{\uparrow}\pm\ket{\downarrow})$.
These sensor states are eigenstates of the signal Hamiltonian with
different eigenvalues which ensures the desired phase difference.
Since this phase difference is what we want to measure we initialize
our qubits to the state: $\frac{1}{\sqrt{2}}(\vert \mathcal{O}_{+} \rangle+\vert \mathcal{O}_{-} \rangle),$ let it acquire a relative phase and measure the probability of the
initial state at the end. 

In this case, $H_{nh}$ reads (in the eigenbasis of $\sigma_{z}$): $H_{nh}=\left(\begin{array}{cc}
-i\gamma & g\\
g & 0
\end{array}\right)$ and the effect of the jump(decay) is $\vert \pm \rangle \rightarrow\frac{1}{\sqrt{2}} \vert \downarrow \rangle$.
In order to restore the oscillations we need to correct
the non Hermitian term in the effective Hamiltonian and the effect
of the jump. The jump is corrected by applying the following unitary operator:
$C= \vert +0\rangle  \langle \downarrow0 \vert +\vert -1\rangle \langle  \downarrow1 \vert + \vert -0\rangle \langle  \uparrow0\vert+\vert+1\rangle \langle  \uparrow1\vert$ upon detecting a photon which could be implemented as: $\exp(-i\frac{\pi}{4}\sigma_{y}^{1})\exp(-i\frac{\pi}{4}\sigma_{z}^{2})\exp(i\frac{\pi}{4}\sigma_{x}^{1}\sigma_{z}^{2})\exp(-i\frac{\pi}{4}\sigma_{x}^{1}).$ It is easy to see that the relative phase between the
sensor states is not harmed by the jump and the subsequent correction.
The dephasing can be corrected by adding the term $\frac{\gamma}{2}\sigma_{y}^{1}\sigma_{z}^{2}$
to the Hamiltonian, which can be seen by observing that this way $\ket {+0}$
will evolve according to $H_{nh}=\left(\begin{array}{cc}
-i\gamma & g+i\frac{\gamma}{2}\\
g-\frac{i\gamma}{2} & 0
\end{array}\right)$ while $\ket{-1}$ will evolve according to $H_{nh}=\left(\begin{array}{cc}
-i\gamma & g-i\frac{\gamma}{2}\\
g+\frac{i\gamma}{2} & 0
\end{array}\right).$  This way both sensor states will be eigenstates of $H_{nh}$ with
eigenvalues $\pm ig-\frac{\gamma}{2}$. It is noteworthy that this
protocol does not require measuring the sensor states but simply a continous
monitoring of the environment. 

There are other ways to correct the
dephasing, but the advantage of the latter is that it is an exact and not
an approximate correction. Two other ways to eliminate the dephasing
are opening energy gaps and using the quantum Zeno effect. Note that the
effect of the dephasing can be seen as mixing the desired
sensor states $\ket {+0},\ket{-1}$ and the undesired states $\ket{-0},\ket {+1}.$
This mixing can be avoided by opening a large energy gap between the
desired and undesired states which could also be achieved by realizing the $\Omega \sigma_x^1 \sigma_z^2$ Hamiltonian. However, the strength of the Hamiltonian must be much larger than the decay rate, i.e., $\Omega \gg \gamma.$ The advantage of this scheme over the previous one is the robustness of the protocol to noise in $\Omega,$ which can be tolerated as long as it is much smaller than $\Omega$ itself.

The other possibility is to use the Zeno effect, i.e., measuring the relevant subspaces
prevents mixing from occurring. The disadvantage of
these two methods is that they provide an approximate correction since
the mixing is eliminated only at the 1$^{st}$ order of $\frac{\gamma}{\Omega}$
, where $\Omega$ is the energy gap or at the 1$^{st}$ order of $\gamma \Delta t,$
the time interval between measurements. The advantage of these methods
is that they do not require knowing $\gamma$ and are more robust
to noise.

\section{Example II} Since many architectures
do not provide ``good'', undecaying qubits, we need to extend our
protocol to the case where all qubits are decay-prone. The extension
is straightforward and involves replacing the good qubit with two
ancilliary bad qubits in the states: $\ket{\uparrow\uparrow\pm\downarrow\downarrow}$. Thus the sensor states
are $\ket{\mathcal{O}_{+}}=\ket{+}\ket{\Psi^{+}}$ and $\ket{\mathcal{O}_{-}}=\ket{-}\ket{\Psi^{-}}$,
with $\ket{\Psi^{\pm}}=\frac{1}{\sqrt{2}}(\ket{\uparrow\uparrow}\pm\ket{\downarrow\downarrow})$.
The ancilliary qubits are initialized in states that can tolerate
up to one decay. Decay in the ancilliary qubits is detected by monitoring
their parity and correcting upon detecting odd parity, and decay of
the sensitive qubit is measured by a click in the detector. Correction
of the dephasing can be done in each of the three methods mentioned above.
Note that in this case precise correction of the dephasing is done
by adding the term: $-\gamma(\sigma_{z}^{1}\sigma_{z}^{2}+0.5)\sigma_{y}^{1}\sigma_{x}^{2}\sigma_{x}^{3}$
to the Hamiltonian.

An analogous protocol works for any signal in $X-Y$ plane: for a signal of the form $g(\cos(\theta)\sigma_{x}+\sin(\theta)\sigma_{y})$ one needs to employ the sensor states $|\uparrow+e^{i\theta}\downarrow\rangle|0\rangle,|\uparrow-e^{i\theta}\downarrow\rangle|1\rangle$ and apply a similar correction. In fact we claim that all single body signals except for $\sigma_{z}$ can be sensed with this scheme. A signal of the form: $H=g(\sin(\theta)\cos(\phi)\sigma_{x}+\sin(\theta)\sin(\phi)\sigma_{y}+\cos(\theta)\sigma_{z}),$ where $\sin(\theta)\neq0$, can be sensed if we eliminate the $\sigma_{z}$ term. This can be performed by opening an appropriate energy gap. Hence if we utilize $|\uparrow+e^{i\phi}\downarrow\rangle|0\rangle,|\uparrow-e^{i\phi}\downarrow\rangle|1\rangle$ as sensor states and apply an energy gap between this subspace and the undesired subspace: $|\uparrow-e^{i\phi}\downarrow\rangle|0\rangle,|\uparrow+e^{i\phi}\downarrow\rangle|1\rangle$ , $H$ can be sensed. We will show later that $\sigma_{z}$ cannot be sensed using a standard photodetection.

These protocols can of course be extended for $N$ qubits and sensing of $H=g\sum_{i=1}^N \sigma_{x}^{i}.$ In this case
one can just use the states $\ket {+}^{N}$ , $\ket {-}^{N}$ as sensor states.
However, this choise requires the use of $N$ photodetectors, one
for each qubit. Note that with $\ket {+}^{N} \ket{0}$, $\ket{-}^{N} \ket{1}$ as sensor
states only one photodetector for the entire system is needed, since
once a photon is detected a measurement of the computational
basis of the sensitive qubits can be made and then the corresponding correction can be applied. 

\section{On the necessity of photodetection} Note that our protocol
will not work using error correction instead of photodetection since
the decayed states are not orthogonal to the undecayed states. In order
to justify the use of photodetection we need to show that it is impossible
to sense single body operators at Heisenberg-Scaling resolution
using the current error-correction scheme. In order to show this, we prove that 
the sensing requirement contradicts the error correction condition. Recall that the sufficient and necessary error correction condition is: 
$\bra{\psi_\alpha} E_i^\dagger E_j \ket{\psi_\beta} = \alpha_{ij} \delta_{\alpha,\beta},$ where $\ket{\psi_{\alpha,\beta}}$ are the code states and $E_{i,j}$ are the errors. In the case of $T_1$ the error operators are $\sigma_-$ and the identity (the original error includes dephasing instead of identity, but since dephasing can be corrected, as explained above, it becomes identity). Hence for $T_{1}$ noise the error correction condition dictates the following conditions: $ \langle 1\vert \sigma_-\vert 1 \rangle =  \langle 2\vert \sigma_- \vert 2 \rangle,$ and $\langle 1\vert \sigma_+\sigma_- \vert1 \rangle = \langle 2\vert \sigma_+\sigma_- \vert 2 \rangle ,$ (where $\ket {1}$ and $\ket {2}$ are the code states) from which one can easily see that $ \langle 1\vert \sigma_{\theta}\vert 1 \rangle =  \langle 2\vert \sigma_{\theta} \vert 2 \rangle,$ to any $\sigma_{\theta}$. This shows that any error correction code resistant to $T_1$ noise that is invariant under $\sigma_{\theta}$ is necessarily an eigenspace of it. Hence in the presence of $T_1$ noise, single body operators cannot be sensed with error correction.      


Thus
photodetection provides the ability to eliminate the decay
in some cases where error-correction does not, but it is limited as
well. The limitation is due to the fact that the jump cannot be corrected for every sensor state. In order to correct the jump the
sensor states must satisfy : $\langle 1 \vert \sigma_+\sigma_- \vert 1 \rangle = \langle 2 \vert \sigma_+\sigma_-\vert 2 \rangle
,$ otherwise every jump will cause deformation of the state. 
In other words only the diagonal part of  the error correction condition is still valid, namely $\bra{\psi_\alpha} E_i^\dagger E_i \ket{\psi_\beta} = \alpha_{i} \delta_{\alpha,\beta}.$
Hence
we get: $\langle 1\vert \sigma_z \vert 1 \rangle =\langle 2 \vert \sigma_z \vert 2 \rangle $ which means that
the photodetection method does not work for any Hamiltonian of the
form $H=\sum a_i\sigma_z^i.$ 

\section{Sensing of $\sigma_{z}$}It may seem strange
that in the presence of $T_1$ noise the only single-body signal that
cannot be sensed with photodetection is $\sigma_{z}$. But note that different kinds of detection provide the ability to sense this operator. Utilizing continous homodyne monitoring is indeed useful\cite{homodyne2000}. Recall that by performing this kind of monitoring the jump operator is changed to $\sigma_{-}+b,$ where $b$ is a real number and the non Hermitian effective Hamiltonian is given by: $H_{nh}=g\sigma_{x}+b\gamma\sigma_{y}-i\gamma\sigma_{+}\sigma_{-}.$ Since $\gamma$ and $b$ are known, the term $b\gamma\sigma_{y}$ can be eliminated such that the resulting $H_{nh}$ will be: $H_{nh}=g\sigma_{x}-i\gamma\sigma_{+}\sigma_{-}$. Now we can apply a similar error correction analysis and observe that in this case the only single body operator that cannot be sensed is $\frac{1}{2}\sigma_{z}+b\sigma_{x}.$ Hence $\sigma_{z}$ can be sensed: since $\sigma_{z}=\frac{1}{(b^{2}+\frac{1}{4})}\left[\frac{1}{2}\left(\frac{\sigma_{z}}{2}+b\sigma_{x}\right)-b\left(-b\sigma_{z}+\frac{\sigma_{x}}{2}\right)\right],$ then we employ a code that is suitable for sensing $\sigma_{\theta}=\frac{1}{\sqrt{b^{2}+\frac{1}{4}}}\left(-b\sigma_{z}+\frac{\sigma_{x}}{2}\right)$, i.e. $\left\{ |\uparrow_{\theta}\rangle|1\rangle\,,\,|\downarrow_{\theta}\rangle|0\rangle\right\}.$ The effect of the term: $\frac{\sigma_{z}}{2}+b\sigma_{x}$ and the dephasing can be eliminated by opening the appropriate energy gap or use Zeno effect. Hence homodyne detection allows sensing of $\sigma_{z}$.

\begin{figure}
\centering
\includegraphics[scale=.4]{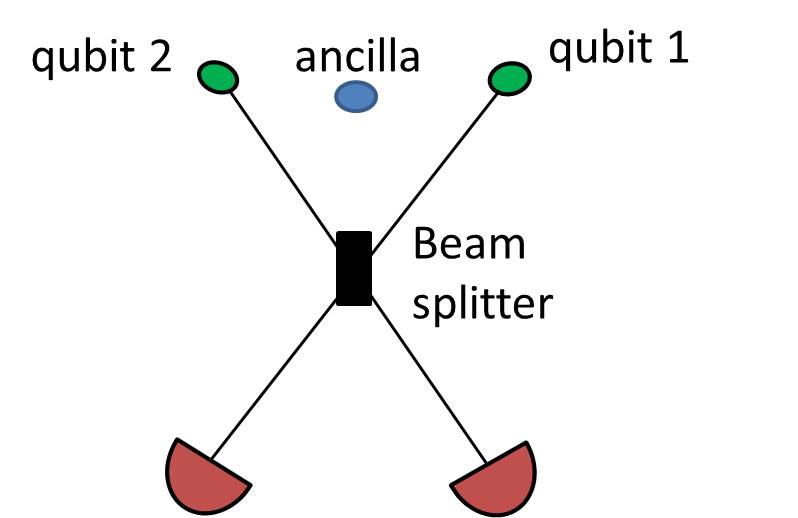}
\caption{ A possible setting for continous monitoring of $\sigma_{-}^{1}+\sigma_{-}^{2}$ , $\sigma_{-}^{1}-\sigma_{-}^{2},$ that provides the ability to sense $\sigma_{z}.$}
\label{sensing_z}
\end{figure}

Another possibility is to use two bad qubits and utilize detection of $\sigma_{-}^{1}+\sigma_{-}^{2}$ , $\sigma_{-}^{1}-\sigma_{-}^{2}.$ This can be achieved for example with the setting presented in fig. \ref{sensing_z}. It can be easily seen from the error correction condition that this detection provides the ability to sense $g_{1}-g_{2}$ where the signal is $H=g_{1}\sigma_{z}^{1}+g_{2}\sigma_{z}^{2}.$ An appropriate code for this scheme is: $\left\{ |\uparrow\downarrow\rangle|1\rangle\,,\,|\downarrow\uparrow\rangle|0\rangle\right\}.$ 
    
\section{Decay and oscillation rate due to imperfect correction}
We now consider the realistic scenario of imperfect correction. Our
scheme works flawlessly under the assumption of perfect gates and
immediate correction. We address the point of non-immediate
correction, by  assuming a constant time interval $\tau$ between the jump and the
correction. In this time interval the decayed sensor states evolve
in an undesired manner according to $H_{nh}.$  Taking this into account
we see that different trajectories yield different behaviors and
averaging over all possible trajectories gives a decay rate and a
shift in the oscillation rate. 

\begin{figure}
\centering
\includegraphics[scale=.2]{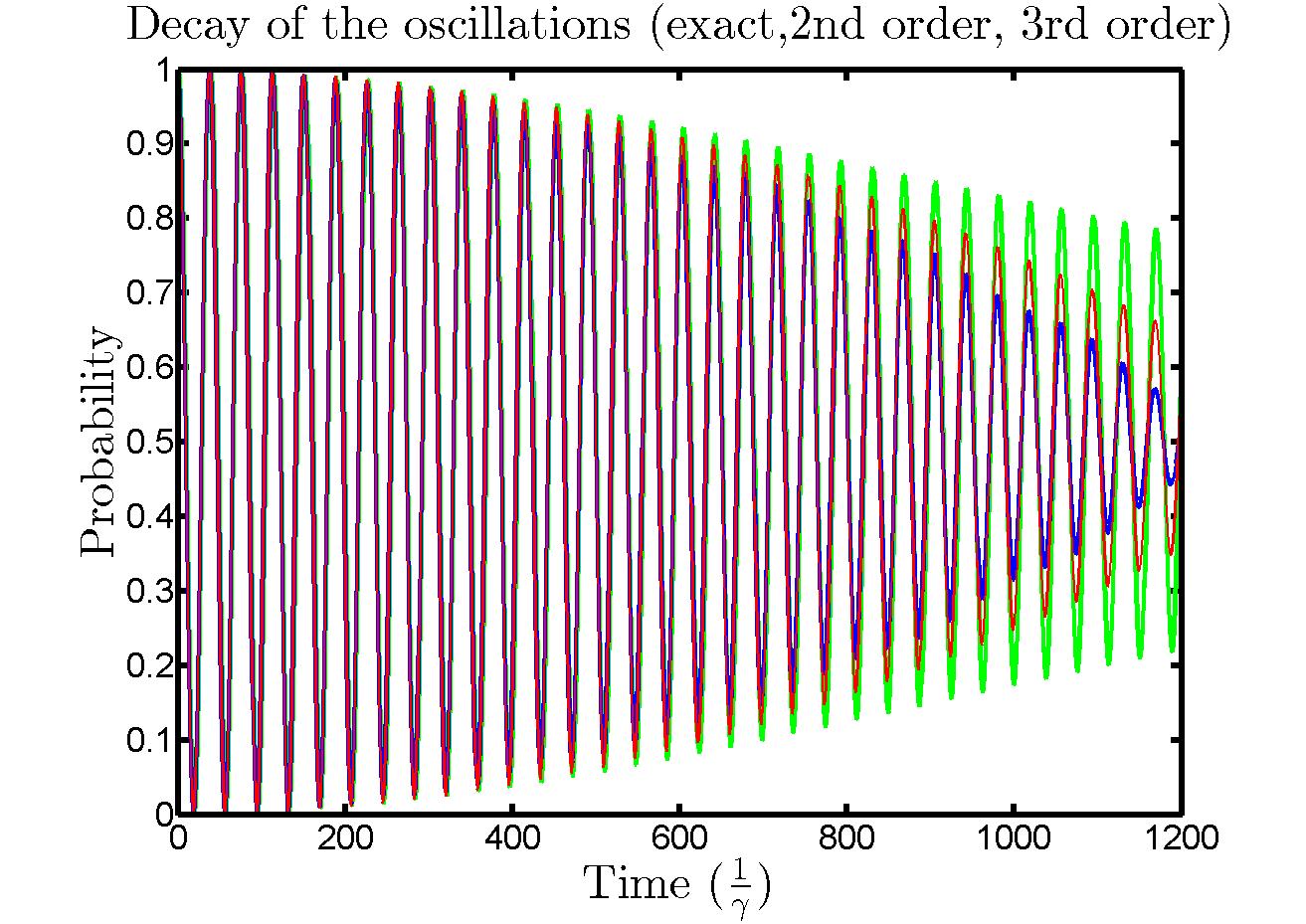}
\caption{The oscillation curve of the approximate result derived in Eq.\ref{eq_approx}, for $\tau=0.2 (\frac{1}{\gamma}),g=0.2 (\gamma), \gamma=1$. 
Green is the full expression, blue corresponds to a correction up to second order(Eq. \ref{eq_approx}) and red corresponds to a correction up to the third order. It can be seen that the derived result(blue) fits the full expression for a short time and sets a lower bound for longer ones.}
\label{Curve_Approx}
\end{figure}

In the following we estimate the shift and the decay for the good
qubit protocol with the exact correction. 
The finite correction time  can be modeled as a delay in the operation of the correction. In this time interval $\exp(-iH_{nh}\tau)$ operates on the states
$ \ket {\downarrow0}$ , $\ket {\downarrow1}.$ As a result, the states $\ket {\uparrow0}$, $\ket {\uparrow1}$
are also populated and immediately corrected to the ``wrong subspace''
$\ket {-0}$, $\ket {+1}$. These states now evolve according to $H_{nh}$; however,
once the next decay occurs they are corrected to the right sensor
states. As a result, the sensor states not only accumulate the
desired phase but also a phase with the wrong, opposite sign. This accounts
for the shift and decay of the oscillations. Since the sequence
of decay, non-Hermitian evolution, correction and non-Hermitian evolution
repeats itself, it can be shown that the projection of the state on
the sensor states is: 
\[
a(t_{1})...a(t_{N})\ket{+0}+b(t_{1})...b(t_{N})\ket{-1},
\]
 where $a(t)=e^{-igt}\cos(g\tau)-i\sin(g\tau)e^{igt}\frac{(2g+i\gamma)^{2}}{4g^{2}}-i\frac{\gamma^{2}}{4g^{2}}\sin(g\tau)e^{-igt}$
, $b(t)=\cos(g\tau)e^{igt}-ie^{-igt}\sin(g\tau)\frac{(2g-i\gamma)^{2}}{4g^{2}}-ie^{igt}\sin(g\tau)\frac{(\gamma^{2}+4ig\gamma)}{4g^{2}}.$
Thus it can be seen that after each sequence the states obtain a phase
in the right direction but also a phase in the wrong direction. 
The probability can be calculated (a detailed derivation can be found in the appendix), by averaging over the different
$t_{i}$ , and assuming $N \gg 1,$ we obtain:
\begin{eqnarray}
p=\frac{1}{2}+\frac{1}{2}\cos(2gt(1-\gamma\tau))-\frac{t(1-\gamma\tau)\tau^{2}g^{3}}{6}&\cdot& \nonumber \\
\sin(2gt(1-\gamma\tau))-\frac{\tau^{2}(t(1-\gamma\tau))^{2}g^{4}}{9}\cos(2gt(1-\gamma\tau)).&&
\label{eq_approx}
\end{eqnarray}
Note that $t$ is shifted to $t(1-\gamma\tau)$ since $\sum_{i}t_{i}=t(1-\gamma\tau)$. Thus we get a decay that goes as $\exp(-\frac{2}{9}(t(1-\gamma\tau))^{2}\tau^{2}g^{4}),$ i.e., decay with rate $\sqrt{\frac{2}{9}}g^2(1-\gamma\tau)\tau,$
and an oscillation rate of $(g+\frac{g^{3}\tau^{2}}{3})(1-\gamma\tau)$. A comparison of this behavior with the full expression is shown in fig. \ref{Curve_Approx}. This result is valid up to a correction that scales as $g^4 \gamma \tau^3 t^2;$ in other words, the approximation is correct for times $\frac{1}{\gamma} \ll t \ll \frac{1}{g^2 \sqrt{\gamma \tau^3}} .$ The lower bound is needed due to the assumption that the number of steps is much larger than one.

\section{Imperfect monitoring}
Non ideal photodetection manifests itself in imperfect quantum efficiency, non vanishing dark counts rate and dead time. All of these can limit the coherence time of the detector and thus its sensitivity. We shalll focus now on the effect of imperfect quantum efficiency, i.e. photon loss with probability $\alpha.$ Qualitatively we would expect the new decay rate to be $\alpha\gamma,$ when
$\gamma$ is the original decay rate. In the case of perfect detection and correction the Master equation is: 
\begin{equation}
\overset{.}{\rho}=-i[H_{nh},\rho]+2\gamma C\sigma_{-}\rho\sigma_{+}C^{\dagger},
\end{equation}
where $C$ is the correction operator. In the case of an imperfect photodetection with
efficiency $1-\alpha$ the new Master equation is: 
\begin{equation}
\overset{.}{\rho}=-i[H_{nh},\rho]+2\gamma((1-\alpha) C\sigma_{-}\rho\sigma_{+}C^{\dagger} + \alpha \sigma_{-}\rho\sigma_{+}).
\end{equation}
We now analyze the solutions of this equation for two correction methods: adding a Hamiltonian term and applying error-correction
of $\sigma_{z}$ errors. By applying the first method we obtain a non-trivial dependence
on $\alpha.$ However for small enough $\alpha$ ($\alpha<0.08$) the probability
is given by $p=\frac{1+cos(m_{1}t)e^{-m_{2}t}}{2},$ where the decay
rate is $m_{2}=2\gamma \alpha+4\gamma \alpha^{2}$ and the oscillation rate
is $m_{1}=2g+2g\alpha-24g\alpha^{2}.$ So for small enough $\alpha$ the sensitivity would be 
$\sqrt{\frac{2\gamma e}{T}}\sqrt{\frac{\alpha}{2}}.$ 
Table .\ref{tab} below shows some exact values.

\begin{table}
\begin{ruledtabular}
\begin{tabular}{|c|c|c|c|c|}
\hline 
$p$ & $0.01$ & $0.03$  & $0.05$  & $0.08$\tabularnewline
\hline 
\hline 
Decay & $0.02\gamma$  & $0.06\gamma$ &  $0.11\gamma$  & $0.2\gamma$\tabularnewline
\hline 
Frequency & $2.02g$ & $2.09g$  & $2.2g$  & $2.66g$\tabularnewline
\hline 
Sensitivity & $\sqrt{\frac{2\gamma e}{T}}0.07$  & $\sqrt{\frac{2\gamma e}{T}}0.11$  & $\sqrt{\frac{2\gamma e}{T}}0.15$ & $\sqrt{\frac{2\gamma e}{T}}0.16$ \tabularnewline
\hline 
\end{tabular}
\end{ruledtabular}
\caption{The exact values for the decay and frequency derived from the full expression.}
\label{tab}
\end{table}

When choosing the second method, i.e., applying error correction of $\sigma_{z}$ errors one gets a much simpler expression. In the limit of
$\tau_{ec}\rightarrow0$ we always stay in the sensor states, so we
should obtain a reduced Master equation which is valid only for the sensor states:
\begin{equation}
EC(-i[H_{nh},\rho])=-i \left[ \left(
\begin{array}{cc}
g-i\frac{\gamma}{2} & 0\\
0 & -g-i\frac{\gamma}{2}
\end{array}
\right),\rho \right]
\end{equation}
 
\begin{equation}
EC((1-\alpha) C\sigma_{-}\rho\sigma_{+}C^{\dagger}+\alpha\sigma_{-}\rho\sigma_{+})=0.5((1-\alpha)\rho+\alpha\sigma_{z}\rho\sigma_{z}),
\end{equation}
 where $EC$ designates the error correction operation.
 This result can be understood intuitively as after an undetected decay occurs the error-correction
takes us back to our code space but with a relative minus sign. Thus
in this basis we get an effective master equation of : 
\begin{equation}
\rho=-ig[\sigma_{z},\rho]+\gamma\alpha(\sigma_{z}\rho\sigma_{z}-\rho),
\end{equation}
 which indeed corresponds to a decay rate of $\gamma\alpha.$

Regarding non vanishing dark counts rate, then the analysis is quite similar to that of photon loss. Given a photon loss with probability $\alpha$ and a dark counts rate of $\kappa$, then by measuring $\sigma_{z}^{1}$ every $\tau_{ec}$ and after every click of the detector and applying the appropriate correction we would obtain the following effective Master equation:  
\begin{equation}
\rho=-ig[\sigma_{z},\rho]+(\gamma\alpha+\kappa/2)(\sigma_{z}\rho\sigma_{z}-\rho),
\end{equation}
which corresponds to a decay rate of $\gamma\alpha+\kappa/2.$ As for a finite dead time of the detector, so this should not limit the scheme as the signal could be turned off (by refocusing) for a time longer than the dead time.

\section{Outlook and Conclusions}  We have presented a method that combines error correction with photodetection for sensing of single body operators. This method provides significantly improved sensitivities with high efficiency photodetectors and attains Heisenberg scaling in the limit of perfect photodetection. This leaves us with the unresolved questions of whether utilizing error correction or error correction with imperfect photodetection can provide Heisenberg scaling as well.    
 
\emph{Acknowledgements---} A. R. acknowledges the support of the Israel Science Foundation(grant no. 039-8823), the support of the European commission (STReP EQUAM Grant Agreement No. 323714), EU Project DIADEMS, the Marie Curie Career Integration Grant (CIG) IonQuanSense(321798), the Niedersachsen-Israeli Research Cooperation Program and DIP program (FO 703/2-1).

\section*{Appendix}
\emph{Derivation of the modified dynamics due to imperfect correction--}We now derive an estimation for the averaged probability when the
correction is not immediate. The time interval between the jump and
the correction is denoted $\tau$ and is assumed to be equal
for all the steps. The evolution between two corrections is given
by: 
\[
\ket{\psi_{i}}=A(t_{i})\ket{\psi_{i-1}}=C\circ U(\tau)\circ\sigma_{-}\circ U(t_{i})\ket{\psi_{i-1}}
\]
where $\ket{\psi_{i-1}}$ is the state after $i-1$ jumps and corrections,
$U(t)=exp(-iH_{nh}t)$ is the non-Hermitian evolution between jumps,
$t_{i}$ is the time between the $i-1$ correction to the $i$ jump
and $C=\ket{+0}\bra{00}+\ket{-1}\bra{01}+\ket{-0}\bra{10}+\ket{+1}\bra{11}$
is the correction operator. We neglected here the possibility of having
a jump in one of the $\tau$ intervals. This is because the probability
of this jump goes as $\gamma^{3}\tau^{3}$ : during this time interval the probability
of the subspace $|\uparrow0\rangle$ , $|\uparrow1\rangle$ goes as $\gamma^{2}\tau^{2},$
 so the probability of a jump goes as $\gamma^{3}\tau^{3}$ and hence
it is negligible. Thus: 
\[
\ket{\psi_{N}}=A(t_{N})...A(t_{1})\ket{\psi_{0}},
\]
where $\ket{\psi_{0}}=\frac{1}{\sqrt{2}}(\ket{+}\ket{0}+\ket{-}\ket{1})$
is our initial state. For convenience we denote: 
\begin{eqnarray*}
U(\tau)\ket{\downarrow0}=f_{1}(\tau)\ket{\downarrow0}+g_{1}(\tau)\ket{\uparrow0}\nonumber \\
U(\tau)\ket{\downarrow1}=f_{2}(\tau)\ket{\downarrow1}+g_{2}(\tau)\ket{\uparrow1}.
\end{eqnarray*}
We now wish to show that: 
\begin{eqnarray*}
\ket{\psi_{N}}=\alpha_{N}(f_{1}(\tau)\ket{+0}+g_{1}(\tau)\ket{-0})\\
+\beta_{N}(f_{2}(\tau)\ket{-1}+g_{2}(\tau)\ket{+1}),
\end{eqnarray*}
where $\alpha_{N}=a(t_{N})...a(t_{2})e^{-igt_{1}}$ and $\beta_{N}=b(t_{N})...b(t_{2})e^{igt_{1}}$
(explicit form of $a$ and $b$ will be found later). This can be
easily proven by induction. Assuming that: 
\begin{eqnarray*}
\ket{\psi_{N}}=\alpha_{N}(f_{1}(\tau)\ket{+0}+g_{1}(\tau)\ket{-0})\\
+\beta_{N}(f_{2}(\tau)\ket{-1}+g_{2}(\tau)\ket{+1}),
\end{eqnarray*}
then the expression for $|\psi_{N+1}\rangle$ reads: 
\begin{eqnarray*}
\ket{\psi_{N+1}}=CU(\tau)\sigma_{-}U(t_{N+1})\Big[\alpha_{N}(f_{1}(\tau)\ket{+0}+g_{1}(\tau)\ket{-0})\\
+\beta_{N}(f_{2}(\tau)\ket{-1}+g_{2}(\tau)\ket{+1})\Big].
\end{eqnarray*}
We now denote for convenience: 
\begin{eqnarray*}
U(t)\ket{-0}=m_{1}(t)\ket{-0}+n_{1}(t)\ket{+0}\nonumber \\
U(t)\ket{+1}=m_{2}(t)\ket{+1}+n_{2}(t)\ket{-1},
\end{eqnarray*}
so further algebra yields: 
\begin{eqnarray*}
\ket{\psi_{N+1}} = CU(\tau)\Big[\alpha_{N}(f_{1}(\tau)e^{-igt_{N+1}}+g_{1}(\tau)m_{1}(t_{N+1})\\
+g_{1}(\tau)n_{1}(t_{N+1}))\ket{\downarrow0}+\beta_{N}(f_{2}(\tau)e^{igt_{N+1}}\\
+ g_{2}(\tau)m_{2}(t_{N+1})+g_{2}(\tau)n_{2}(t_{N+1}))\ket{\downarrow1}\Big].
\end{eqnarray*}
Now it is straightforward to see that the desired form is obtained, where:
\begin{eqnarray*}
&\alpha_{N+1}=\alpha_{N}(f_{1}(\tau)e^{-igt_{N+1}}+g_{1}(\tau)m_{1}(t_{N+1})\\
&+g_{1}(\tau)n_{1}(t_{N+1})),\\
&\beta_{N+1}=\beta_{N}(f_{2}(\tau)e^{igt_{N+1}}+g_{2}(\tau)m_{2}(t_{N+1})\\
&+g_{2}(\tau)n_{2}(t_{N+1})).
\end{eqnarray*}
In order to obtain the exact form of $\alpha_{N}$ and $\beta_{N}$ we need
to find the initial $\alpha_{1}$ and $\beta_{1}.$ Note that $|\psi_{1}\rangle=e^{-igt_{1}}(f_{1}(\tau)\ket{+0}+g_{1}(\tau)\ket{-0})+e^{igt_{1}}(f_{2}(\tau)\ket{-1}+g_{2}(\tau)\ket{+1}),$
so: 
\[
\alpha_{N}=a(t_{N})...a(t_{2})e^{-igt_{1}}\:,\:\beta_{N}=b(t_{N})...b(t_{2})e^{igt_{1}},
\]
but since $N\gg1$ this can be simplified to: 
\[
\alpha_{N}=a(t_{N})...a(t_{1})\:,\:\beta_{N}=b(t_{N})...b(t_{1}),
\]
where as we have found: $a(t)=f_{1}(\tau)e^{-igt}+g_{1}(\tau)m_{1}(t)+g_{1}(\tau)n_{1}(t)$
, $b(t)=f_{2}(\tau)e^{igt}+g_{2}(\tau)m_{2}(t)+g_{2}(\tau)n_{2}(t).$
 The exact expressions are: 
\begin{eqnarray*}
&a(t)=\frac{1}{4g^{2}}e^{-igt}\Big(4g^{2}\cos(g\tau)\\
&-i\sin(g\tau)(\gamma^{2}+e^{2igt}(2g+i\gamma)^{2})\Big),\\
&b(t)=\frac{1}{2g^{2}}e^{-ig\tau}\Big(2g^{2}\cos(gt)+e^{ig\tau}(2ig^{2}\cos(g\tau)\\
&+(-2g^{2}+4ig\gamma+\gamma^{2})\sin(g\tau))\sin(gt)\Big).
\end{eqnarray*}
We are now reminded that our goal is to calculate the probability
of the initial state $\ket{\psi_{0}},$ but this is just: 
\[
p=\frac{1}{2}+\frac{1}{2}\frac{f_{1}^{*}(\tau)f_{2}(\tau)\alpha_{N}^{*}\beta_{N}+\alpha_{N}\beta_{N}^{*}f_{1}(\tau)f_{2}^{*}(\tau)}{|f_{1}\alpha|^{2}+|f_{2}\beta|^{2}}.
\]
We have omitted from the denominator the probabilities for $\ket{-}\ket{0}$
, $\ket{+}\ket{1}$ . That is because their probabilities are $|\alpha_{N}g_{1}(\tau)|^{2}$
and $|\beta_{N}g_{2}(\tau)|^{2}$ which is just $|g_{1}(\tau)|^{2}$
and $|g_{2}(\tau)|^{2}$ (in leading orders of $\tau^{2}$) . So these
probabilities are of order $g^{2}\tau^{2}$ , $\gamma^{2}\tau^{2}$ and since we have only two
such terms this accounts to a change of order $\tau^{2}$ and not
$N\tau^{2}$ or $N^{2}\tau^{2}$ and thus it is negligible. Plugging
in our expressions for $\alpha_{N}$ , $\beta_{N}$ and since $f_{1}(\tau)=f_{2}(\tau)$
we get: 
\[
p=\frac{1}{2}+\frac{1}{2}\frac{\underset{i}{(\prod}a(t_{i})^{*}b(t_{i}))+h.c.}{\underset{i}{\prod}a(t_{i})^{*}a(t_{i})+\underset{i}{\prod}b(t_{i})^{*}b(t_{i})}.
\]
Now using Mathematica we expand $p$ to leading orders in $\tau$ (assuming
$g\tau,\gamma\tau\ll1$ ), and work in the interesting limit of
$g\ll\gamma$. Thus we can use the approximation $\langle e^{igt_{i}}\rangle\approx e^{i\frac{g}{\gamma}}$
(which is true at 1st order in $\frac{g}{\gamma}$ ). Taking now the leading order of $\frac{g}{\gamma}$ of the full expression we obtain:
\[
p=\frac{1}{2}+\frac{1}{2}\cdot cos(2gT)-\frac{t\tau^{2}}{6}g^{3}sin(2gT)-\frac{1}{9}T^{2}\tau^{2}g^{4}cos(2gT)
\]
Where we denoted $T=\sum_{i}t_{i}$ . Note that $T$ is not the total
time, since the total time includes also the $\tau$ intervals.
So the total time is $t=T+\tau\gamma t$ , hence $T=t(1-\gamma\tau).$
Then the probability is: 
\begin{eqnarray*}
p=&\frac{1}{2}+\frac{1}{2}\cos(2gt(1-\gamma\tau))-\frac{t(1-\gamma\tau)\tau^{2}g^{3}}{6}\sin(2gt(1-\gamma\tau))\\
&-\frac{\tau^{2}(t(1-\gamma\tau))^{2}g^{4}}{9}\cos(2gt(1-\gamma\tau)).
\end{eqnarray*}
The term $-\frac{1}{9}(t(1-\gamma\tau))^{2}\tau^{2}g^{4}\cos(2gt(1-\gamma\tau))$
corresponds to a decay of the oscillations, which goes like $\exp(-\frac{2}{9}(t(1-\gamma\tau))^{2}\tau^{2}g^{4}),$
while the term $-\frac{t(1-\gamma\tau)\tau^{2}}{6}g^{3}\sin(2gt(1-\gamma\tau))$
corresponds to a shift in the oscillation rate yielding an effective
oscillation rate of $(g+\frac{g^{3}\tau^{2}}{6})(1-\gamma\tau)$.

 {}

\end{document}